\documentclass[lettersize,journal]{IEEEtran}
\usepackage{graphicx} 
\usepackage{cite}
\usepackage{amsmath,amssymb,amsfonts}
\usepackage{algorithmicx}
\usepackage{subfigure}
\usepackage{textcomp}
\usepackage{amsthm}
\usepackage{dblfloatfix}
\usepackage{array}
\usepackage{balance}
\usepackage{comment}
\usepackage{hyperref}
\usepackage{soul} 
\usepackage{mathtools}
\usepackage{tikz}
\usepackage{pgfplots}
\usepackage{tabularx}
\usepackage{multirow}
\usepackage{multicol}
\usetikzlibrary{positioning}
\usepackage{cleveref}
\usepackage{orcidlink}
\usepackage{multirow, hhline}
\newcolumntype{Y}{>{\centering\arraybackslash}X}

\Crefname{equation}{Eq.}{Eqs.}
\Crefname{section}{Sec.}{Secs.}
\Crefname{figure}{Fig.}{Figs.}

\def\D{\mathrm{d}}
\newcommand{\Dc}{\mathcal{D}}
\newcommand{\Rc}{\mathcal{R}}

\title{XR Offloading Across Multiple Time Scales:\\The Roles of Power, Temperature, and Energy}
\author{Francesco~Malandrino,~\IEEEmembership{Senior~Member,~IEEE,}
Olga~Chukhno,~\IEEEmembership{Member,~IEEE,}
Alessandro~Catania, \IEEEmembership{Senior~Member,~IEEE,}
Antonella~Molinaro,~\IEEEmembership{Senior~Member,~IEEE,}
Carla~Fabiana~Chiasserini,~\IEEEmembership{Fellow,~IEEE}
\thanks{F.~Malandrino (francesco.malandrino@cnr.it) and C.~F.~Chiasserini (carla.chiasserini@polito.it) are with CNR-IEIIT, Italy. C.~F.~Chiasserini is with Politecnico di Torino and CNIT, Italy. O.~Chukhno (olga.chukhno@unirc.it) and A.~Molinaro (antonella.molinaro@unirc.it) are with Mediterranea University of Reggio Calabria and CNIT, Italy. A.~Catania (alessandro.catania@unipi.it) is with  University of Pisa, Italy. }
\thanks{This work was supported by the innovation programme under Grand Agreement No.\,101095363 and by the European Union under the Italian National Recovery and Resilience Plan (NRRP) of NextGenerationEU (PE00000001 -- program ``RESTART'').}
}

\begin{document}

\maketitle
\begin{abstract}
Extended reality (XR) devices, commonly known as {\em wearables}, must handle significant computational loads under tight latency constraints. To meet these demands, they rely on a combination of on-device processing and edge offloading. This letter focuses on offloading strategies for wearables
and assesses the impact of offloading decisions over three distinct time scales:
instantaneous power consumption, short-term temperature fluctuations, and long-term battery duration. We introduce a comprehensive system model that captures these temporal dynamics, and propose a stochastic and stationary offloading strategy, called TAO (for \textit{temperature-aware offloading}), designed to minimize the offloading cost while adhering to power, thermal, and energy constraints. Our performance evaluation, leveraging COMSOL models of real-world wearables,
confirms that TAO successfully avoids exceeding temperature limits while keeping additional edge offloading to a minimum. These results also highlight how properly accounting for all features of wearables allows fully exploiting edge offloading opportunities.
\end{abstract}



\section{Introduction}\label{sec:intro}

Extended Reality (XR) 
is among the most promising yet demanding applications for next-generation networks, including 6G-and-beyond. 
A key challenge lies in the conflict between the high computational demands of XR services  -- which entail~\cite{ericsson2021xr} such complex tasks as rendering complex virtual environments, processing sensor data, and generating personalized content -- and the strict constraints imposed by the compact form factor of wearable devices. Notably, wearables inherently have limited processing power, constrained battery life, and are susceptible to overheating during intensive on-device computation. Thermal concerns are particularly critical due to the close physical contact between the device and the user's skin: 
excessive heating leads to discomfort and degraded experience while risking low-temperature burns.

\begin{figure}[!t]
    \center{\includegraphics[width=.7\columnwidth]{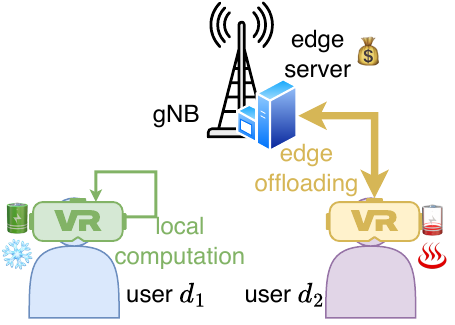}}
    \caption{Combining edge offloading and on-device computation for XR: offloading saves energy at high temperature/low battery (user~$d_2$); local computation reduces costs when conditions are sufficient (user~$d_1$).}\vspace{-1mm}
    \label{fig:main}
\end{figure}

Edge offloading, as exemplified in Fig.~\ref{fig:main}, has long been identified as an effective strategy to mitigate the limitations of XR wearables. By offloading part of the computational workload to an edge server, typically accessed via a cellular base station (eNB in~5G, gNB in~6G), wearables can achieve enhanced performance and reduced local power consumption. This approach has been explored in~\cite{sampath2024enabling} for 5G~networks and in~\cite{cai2024mobility} for 6G, considering 
user mobility and dynamic network conditions. Similarly, the interplay between content dynamics and network resources in edge-aided XR services has been explored in~\cite{chukhno2022interplay,chukhno2023user}. However, offloading is not without drawbacks, since it introduces additional network-related challenges (e.g., signal blockage), increases dependence on edge infrastructure, and incurs communication and computation costs. These trade-offs between local computation and edge offloading have been acknowledged, 
with Ericsson proposing several offloading {\em levels}~\cite{ericsson2023xr}, and Qualcomm leveraging edge resources to augment wearables' capabilities~\cite{qualcomm2025}.

In the edge offloading literature~\cite{feng2022cost,trinh2022deep}, a common goal is to minimize the incurred cost, which means maximizing the amount of computation performed locally. Pursuing this goal requires accurately modeling and properly evaluating {\em all} effects of local computation across multiple time scales:
\begin{itemize}
    \item \textit{short-term}, the (instantaneous) power required for local processing must remain within the device's thermal design power (TDP);
    \item \textit{medium-term}, the device's temperature must stay below $43$~°C, the threshold beyond which low-temperature burns may occur~\cite{Rupp2024};
    \item \textit{long-term}, the energy consumption must not exceed the available battery capacity over the period of interest.
\end{itemize}

Wearables are unique among other mobile devices in that {\em all} the three time scales are equally relevant. While for other devices a metric may raise more concerns than the others, e.g., battery life for smartphones or power for laptops, XR wearables must concurrently respect all metrics to ensure user safety and sustained operation. This multi-timescale sensitivity necessitates more sophisticated offloading decisions. 
However, most of the state-of-the-art approaches fail to properly and jointly account for all relevant time scales,  which results in suboptimal -- and potentially infeasible -- offloading decisions. Similarly, many works focus on average performance, e.g., the expected values of target metrics; however, this is often not  adequate for demanding services like XR.

In this paper, we tackle the challenge of edge offloading for XR wearables with the goal of {\em (i)} minimizing task offloading
, while {\em (ii)} satisfying power, thermal, and energy constraints with a tunable level of confidence. To this end, Sec.~\ref{sec:deterministic} presents a system model capturing how offloading decisions affect power consumption, temperature, and battery level of XR wearables. Sec.~\ref{sec:stationary} introduces our solution TAO (Temperature-Aware Offloading), leveraging our model to make high-quality offloading decisions under uncertain, stochastic demand for XR services. Sec.~\ref{sec:peva} presents a performance evaluation of TAO, comparing it against state-of-the-art methods. 

\section{System Model and Problem Formulation}\label{sec:deterministic}

This section describes the system architecture and problem formulation for XR task offloading. For simplicity, we initially assume that the computational load is known in advance; this assumption is relaxed in Sec.~\ref{sec:stationary}.

{\bf System architecture and requests.}
We consider a system as illustrated in Fig.~\ref{fig:main}, where a set of users~$\Dc=\{d\}$, each equipped with a wearable device, access XR services with optional support from edge servers. XR services require multiple processing tasks~\cite{ericsson2021xr}, including object detection, map optimization, and multimedia processing. We abstract these tasks as a set of processing requests~$r\in\Rc$. Each request~$r$ is generated by a user~$d(r){\in}\Dc$ at time~$0{\leq} a(r){\leq} T$, where~$T$ denotes the length of the time horizon under consideration.
For simplicity, we consider requests to be uniform, with the same size and computational requirements. This assumption can be dropped by adding parameters to the system model, and our solution concept does not depend upon it.

Each request~$r{\in}\Rc$ can be executed either locally on the device or {\em offloaded} to an edge server. This decision is captured by a binary variable~$\ell(r){\in}\{0,1\}$, taking on 1 when the request $r$ is served locally, and 0 when it is offloaded. Our system model captures the trade-offs associated with these decisions:
 {\em (i)}  Local execution increases power consumption, raising device temperature and draining battery;
   {\em (ii)} Offloading reduces local resource usage but incurs communication costs.
The resources devoted to each request remain the same whether offloaded or not: this reflects the fact that we do not seek trade-offs between quality of service (QoS) and costs; rather, we seek to honor all QoS constraints at the minimum cost.

{\bf Local computations and power.}
To model the impact of local computation, we quantify how offloading decisions~$\ell(r)$ influence the device's power. Let~$\delta(r){>}0$ denote the duration required to execute request~$r$, and~$\pi(d,r)$ represent the additional\footnote{Computed with respect to not serving that request locally.} power for serving that request, then power~$p(d,t)$ at which device~$d$ operates at time~$t$ can be expressed as:
\begin{equation}
\label{eq:power}
p(d,t){=}\!\!\!\!\!\!\!\!\!\!\!\!\!{\sum_{\substack{r{\in}\Rc\colon d(r)=d\\ \wedge a(r){\leq} t{\leq} a(r){+}\delta(r)}}}\!\!\!\!\!\!\!\!\!\!\!\!\ell(r)\pi(d,r),\quad\forall d{\in}\Dc,0{\leq} t{\leq} T.
\end{equation}
While nonlinear power interactions may occur under high parallel multi-tasking loads, the linear power accumulation assumption is a validated simplification in thermal modeling of portable electronics~\cite {Lee2008,liu2023dynamic}, such as XR wearables.

{\bf Temperature and other constraints.}
Using the power expression in (\ref{eq:power}), we can now model how offloading decisions impact the three critical operational constraints of XR wearables:
\begin{itemize}
    \item power, which must remain below the device's TDP;
    \item total energy consumption over the time horizon~[$0$,$T$], which must not exceed the device's battery capacity;
    \item temperature, which shall not exceed the $43$~°C threshold.
\end{itemize}

To ensure the TDP is not exceeded at any time, we enforce a constraint on the maximum allowable power~$P^{\max}(d)$: 
\begin{equation}
\label{eq:p-max}
p(d,t){\leq} P^{\max}(d),\quad\forall d{\in}\Dc,0{\leq} t{\leq} T.
\end{equation} 

For battery constraints, let $b(d,t)$ denote the battery level of device $d$ at time $t$, and $b(d,0)$ its (known) initial level; then, we can write the battery level at any time $t$ as:
\begin{equation}
\label{eq:honor-b}
b(d,t){=}b(d,0){-}\!\!\!\int_0^t \!\!p(d,t)\D t,\quad\forall d{\in}\Dc,0{\leq}  t{\leq} T,
\end{equation}
We require that the device does not fully deplete its battery by the end of the time horizon~$T$:
\begin{equation}
\label{eq:b-min}
b(d,T){\geq}0,\quad\forall d{\in}\Dc.
\end{equation}

To capture the relationship between power and temperature, we model the thermal behavior of wearables as a Linear Time-Invariant (LTI) system, consistent with~\cite{liu2023dynamic}: the input of the LTI system is the power~$p(d,t)$, and its output is the temperature~$\tau(d,t)$.
Assuming constant heat capacity and thermal resistance is a commonly adopted assumption in the thermal modeling of XR wearable devices~\cite{Lee2008,Matsuhashi2020}.
Each device has an {\em impulse response}~$h(d,\overline{t})$ and the temperature is given by the convolution of power with the impulse response. The temperature must never exceed its limit~$\tau^{\max}$:
\begin{equation}
\label{eq:tau-max}
~\tau(d,t){=}p(d,t)\ast h(d,\overline{t}){\leq}\tau^{\max}, \quad\forall d{\in}\Dc,0{\leq} t{\leq} T.
\end{equation}

{\bf Objective.}
As discussed earlier and noted in~\cite{trinh2022deep}, edge offloading increases the cost for XR service provisioning; hence, the system aims to use local processing whenever possible. Given that all requests must be served, and subject to constraints (\ref{eq:p-max}--\ref{eq:tau-max}), the objective becomes maximizing the number of locally-served requests, i.e., 
$\max_{\ell}\sum_{r{\in}\Rc}\ell(r)$.

\section{The TAO solution strategy}
\label{sec:stationary}

We now drop the assumption of known request arrival times and present TAO, a strategy designed to operate under uncertainty. TAO is characterized by the following key features: \begin{itemize}
    \item {\em stochastic}: 
    TAO uses 
    a probability~$\alpha_r{\in} [0, 1]$ that an incoming request is served locally;
    \item {\em stationary}: the probability $\alpha_r$~value is time-invariant; 
    \item {\em uncertainty-aware}: TAO guarantees that all system constraints are met with at least a given confidence level~$\omega$.
\end{itemize}

Requests arrive at each device~$d$ according to a Poisson process with rate~$\lambda_d$. Accordingly, $N_d(T)$ is a random variable with expected value~$\lambda_d \alpha_d T$, denoting the number of locally served requests at device $d$ within $[0, T]$. Given the local service probability~$\alpha_r$, decision variables $\ell(r)$ become Bernoulli random variables with parameter $\alpha_r$, such that $P(\ell(r){=}1){=}\alpha_r$ and $P(\ell(r){=}0){=}1{-}\alpha_r$. The objective remains to maximize the number of locally served requests, while satisfying the constraints on the power, battery level, and temperature as follows:
\begin{equation}
\label{eq:Stochastic_obj}
\max_{\alpha} \sum_{r {\in} \mathcal{R}} \alpha_r \quad\text{subject to:}
\end{equation}
\begin{equation}
\label{eq:Stochastic_power_sufficiency}
P^{\max}(d) {\geq} p(d,t) {\geq} \!\!\!\sum_{r: d(r){=}d}\!\!\!\! \pi(d,r) \ell(r), \quad\forall d{\in}\Dc,0{\leq} t{\leq} T,
\end{equation}
\begin{equation}
\label{eq:Stochastic_battery}
b(d,t) {>} 0, \!\!\quad\forall d{\in}\Dc,
\text{ and }
\tau(d,t) {\leq} \tau^{\max}, \!\!\quad\forall d{\in}\Dc,0{\leq} t{\leq} T,
\end{equation}
where $p(d,t)$, $b(d,t)$, and $\tau(d,t)$ are the random variables representing, respectively, the power consumption,  battery level, and temperature of device $d$ at time $t$.

Solving this stochastic optimization problem requires deriving the cumulative distribution functions (CDFs) of $p(d,t)$, $b(d,t)$, and $\tau(d,t)$ as follows:
\begin{equation}
F_{p/b/\tau(d,t)}(x) {=} \int_{0}^{x} f_{p/b/\tau(d,t)}(u) \, du\,,
\end{equation}
where $f_{p/b/\tau(d,t)}$ represents the probability density functions (PDFs) of the power $p(d,t)$, battery level $b(d,t)$, and temperature $\tau(d,t)$ at time $t$.

We can formulate the robust constraints, that ensure the system remains within safe operational bounds for power, temperature, and battery usage with a specified  confidence level~$\omega{\in}(0, 1)$. To this end, we exploit the fact that the CDF evaluated at~$\hat{t}$ yields the probability of the random variable being less than or equal to~$\hat{t}$, which allows us to write:
\begin{equation}
F_{p(d,t)}(T_p) {\leq}\ \omega, \
F_{\tau(d,t)}(T_\tau) {\leq}\ \omega, \ \text{and} \
F_{b(d,t)}(T_b) {\leq}\ \omega,
\end{equation}
where $T_p$, $T_\tau$, and $T_b$ are the $\omega$-quantiles of $p(d,t)$, $\tau(d,t)$, and $b(d,t)$, respectively.

By exploiting CDFs instead of average values, our formulation provides robust guarantees against critical events, such as power overconsumption, battery depletion, and overheating. Our methodology remains fully applicable even when explicit, closed-form expressions for input functions (e.g., the impulse response~$h(d,\bar{t})$) are unavailable and must be obtained through measurements. However, when closed-form expressions are available, TAO can compute the exact optimal offloading probability, as demonstrated next.

\subsection{A closed-form example}

We derive the $\omega$-quantile constrained CDFs and their deterministic equivalents for specific PDFs of $p(d,t)$, $\tau(d,t)$, and $b(d,t)$.
Recalling that request arrivals at each device $d$ follow a Poisson process with rate $\lambda_d$, the power increase for all requests at device $d$ is constant and given by $\pi(d,r){=}\pi_d$. Accordingly, the number of locally served requests $N_d(t)$ is a Poisson random variable with mean $\lambda_d \alpha_d t$, and the power consumption at device $d$ at time $t$ is directly proportional to $N_d(t)$ and to the power increase per request $\pi_d$: 
\begin{equation}
p(d,t){=}\pi_d N_d(t).
\end{equation} 
As the maximum power $P^{\max}(d)$ is also constant per device~$d$, the probability that the power consumption of device $d$ does not exceed $P^{\max}(d)$ at time $t$ can be rewritten as:
\begin{gather}\label{eq:11}
\Pr(p(d,t){\leq}P^{\max}(d)){=}\Pr(N_d(t){\leq}\frac{P^{\max}(d)}{\pi_d}),
\end{gather}
which, using the CDF of the Poisson distribution, can be defined as follows:
\begin{gather}
\Pr(p(d,t){\leq}P^{\max}(d)){=} \!\!\!\!\!\sum_{i{=}0}^{\lfloor \! \frac{P^{\max}(d)}{\pi_d} \rfloor} \frac{(\lambda_d\alpha_d t)^i e^{{-}\lambda_d\alpha_d t}}{i!}.
\end{gather} 
By incorporating the $\omega$-quantile, we obtain:
\begin{equation}
\sum_{i{=}0}^{\lfloor \frac{T_p}{\pi_d} \rfloor} \! \frac{(\lambda_d\alpha_d t)^i e^{{-}\lambda_d\alpha_d t}}{i!}{\leq}\omega.
\end{equation}

Therefore, the constraint ensuring that device $d$ has sufficient power to serve  $N_d(T)$ requests is given~by:
\begin{gather}
\label{Stochastic_power_sufficiency_pr}
\Pr({\pi_d}N_d(t) {\geq} {\pi_d}N_d(T))
{=} \Pr(N_d(t) {\geq} N_d(T)){=}\nonumber \\ \sum_{i{=}0}^{\infty} \sum_{j{=}i}^{\infty} \Pr(N_d(t){=}j)~ \Pr(N_d(T){=}i).
\end{gather}
It follows that:
\begin{gather}
\sum_{i{=}0}^{\infty} \sum_{j{=}i}^{\infty} \frac{(\lambda_d\alpha_d t)^j e^{-\lambda_d\alpha_d t}}{j!} \frac{(\lambda_d\alpha_d T)^i e^{-\lambda_d\alpha_d T}}{i!}{\leq}\omega.
\end{gather}

We then require the device battery to maintain a non-negative level at time $t{=}T$. Given the power consumption $p(d,t){=}\pi_d N_d(T)$, the battery level $b(d,t)$ can be expressed~as:
\begin{gather}
b(d,t) {=} b(d,0) {-} \pi_d \int_0^t N_d(t) \, dt {=} \nonumber\\
b(d,0) {-} \pi_d \lambda_d\alpha_d \int_0^t t \, dt {=} b(d,0) {-} \pi_d \lambda_d\alpha_d \frac{t^2}{2}.
\end{gather} 
To ensure a positive battery level at time $t{=}T$, we have:
\begin{equation}
b(d,T) {=} b(d,0) {-} \pi_d \lambda_d\alpha_d \frac{T^2}{2} {>} 0.
\end{equation} 
Thus, we obtain
\begin{equation}
\sum_{i{=}0}^{\lfloor T_b/(b(d,0) {-} \pi_d \lambda_d\alpha_d \frac{T^2}{2}) \rfloor} \frac{(\lambda_d\alpha_d T)^i e^{{-}\lambda_d\alpha_d T}}{i!} {\leq}\omega.
\end{equation}

Finally, we obtain a closed-form expression for the impulse response~$h(d,\bar{t})$. A typical form of such a function is represented by a series of two first-order responses~\cite{billings2010}. This involves an initial negative exponential response approaching a certain high value, followed by a second negative exponential response converging towards a lower value. The impulse stimulus, characterized by an amplitude $A$ over the interval~$[0, T]$, can be decomposed into the superposition of two step functions: $A u(t)$ and ${-}A u(t{-}T)$. Consequently, the output of the linear system, denoted as $\tau(d,t)$, can be represented as the superposition of these two responses:
\begin{equation}
\tau(d,t){=} B \left[ (1 {-} e(-t/\theta)) {-} (1 {-} e({-}(t{-}T)/\theta)) \right],
\end{equation}
where $B$ is the asymptotic output for a step input of $A u(t)$, $B=A R_{th}$, where $R_{th}$ (in °C/W) is the equivalent thermal resistance of the system, which indicates the temperature increase for each unit of dissipated power, and $\theta$ is the time constant. 

The $\omega$-quantile constrained CDF is then given by:
\begin{equation}
\!\!\!\!\sum_{i{=}0}^{\lfloor T_\tau/B \left[(1{-}e({-}t/\theta)){-}(1{-}e({-}(t{-}T)/\theta)) \right] \rfloor}\!\!\frac{(\lambda_d\alpha_d t)^i e^{-\lambda_d\alpha_d t}}{i!}{\leq}\omega.
\end{equation}

\section{Performance evaluation}
\label{sec:peva}

Our performance evaluation focuses on two commercially available XR wearables: \textit{Microsoft's HoloLens}, powered by the Intel Cherry Trail System-on-chip (SoC), and \textit{Google Glass}, equipped with a Texas Instrument's OMAP~4430 processor. Key characteristics relevant to our study are summarized in Table~\ref{tab:hw}. For modeling purposes, we assume that both devices consume negligible power in idle states (i.e., when not performing local computations), and operate at their TDP when executing local processing tasks\footnote{More advanced features like thermal throttling could be seamlessly embedded in our model, at the price of additional complexity.}. Notably, due to its more capable processor, the HoloLens completes request processing faster than Google Glass.

\begin{figure}[!t]
    \center{\includegraphics[width=0.99\linewidth]{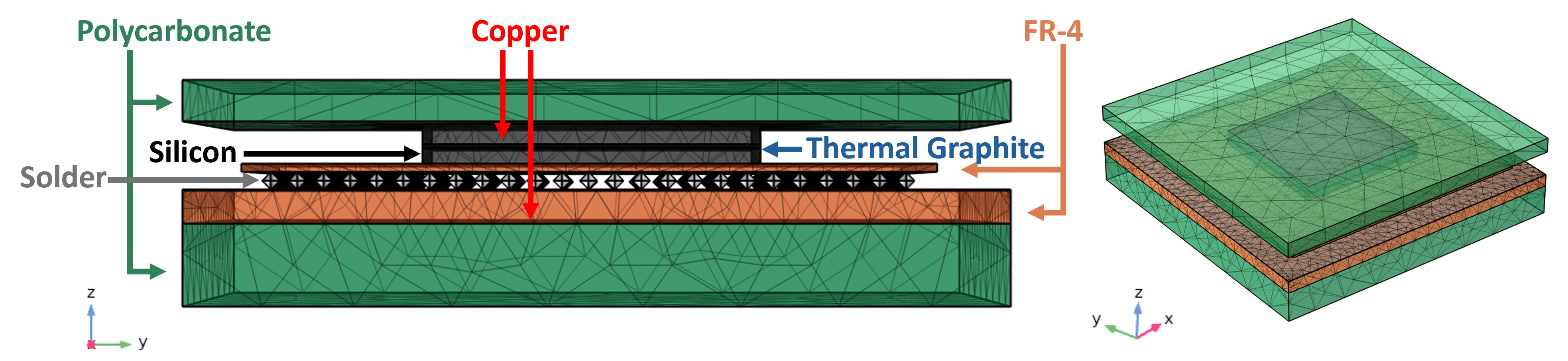}}
    \caption{Lateral view (left) and isometric view (right) of the 3D structure simulated in COMSOL for a chip used in wearables (Intel Cherry Trail, used in Microsoft HoloLens, in this case).}
    \label{fig:3d-model}
\end{figure}

To analyze the thermal behavior of the selected wearables, we use the COMSOL software to simulate 3D models of each device, such as the Cherry Trail SoC presented in Fig.~\ref{fig:3d-model}. It includes a polycarbonate outer shell, a printed circuit board (PCB) substrate, copper interconnections, solder layers with a ball grid array, a package substrate, a silicon die, thermal graphite as a thermal interface material, a copper heat sink, and an inner polycarbonate shell. The heat source is uniformly distributed in the silicon die, and external boundary conditions are defined by a heat transfer coefficient of $8.3 \mathrm{W/(m^2K)}$, simulating still air~\cite{cengel2015}. 

\begin{figure*}
\centering
\includegraphics[width=.3\textwidth]{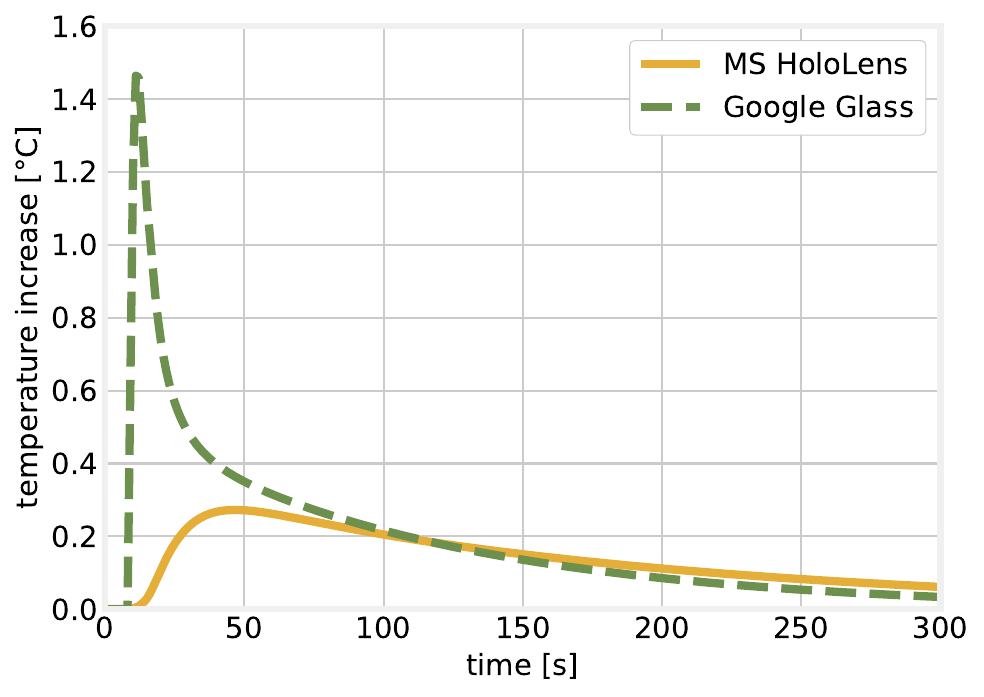}
\includegraphics[width=.3\textwidth]{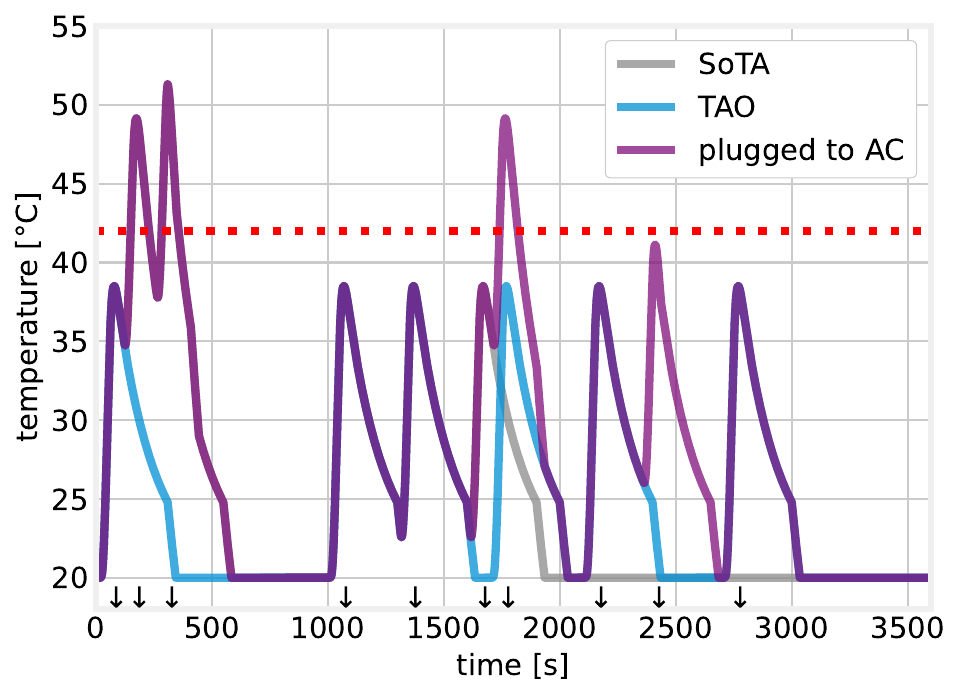}
\includegraphics[width=.3\textwidth]{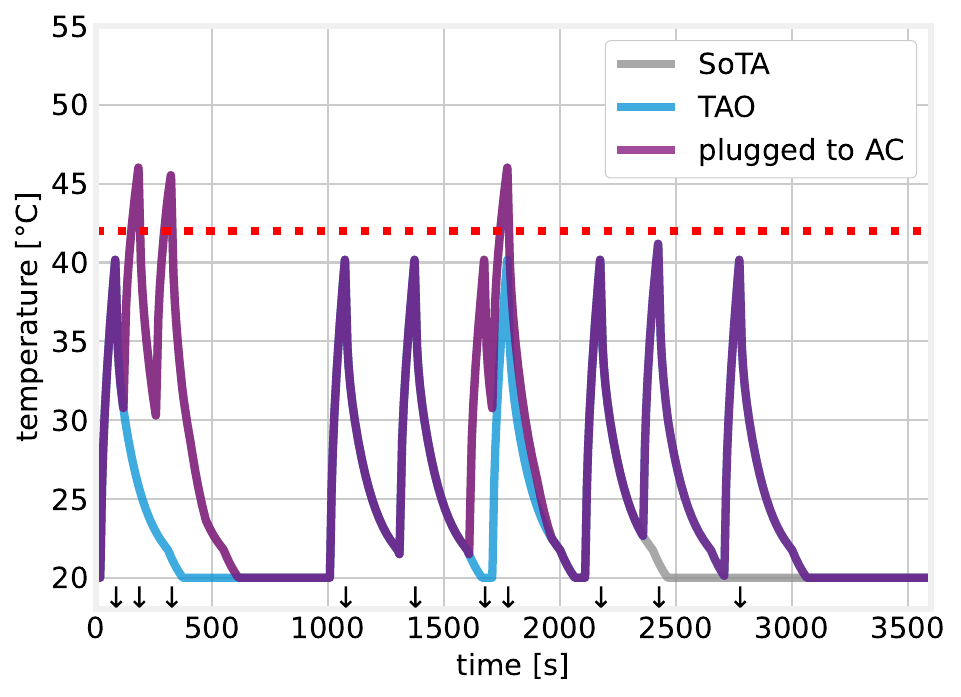}
\caption{
Impulse responses for the considered wearables (left); time evolution of the temperature for Microsoft HoloLens (center) and Google Glass (right). 
\label{fig:peva1}
}
\includegraphics[width=.30\textwidth]{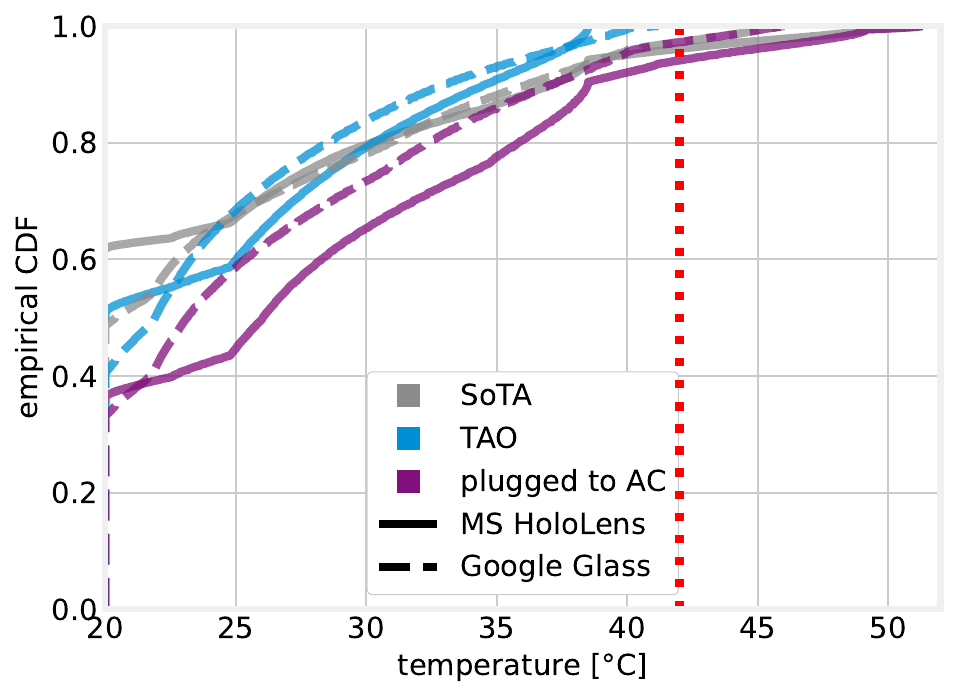}
\includegraphics[width=.30\textwidth]{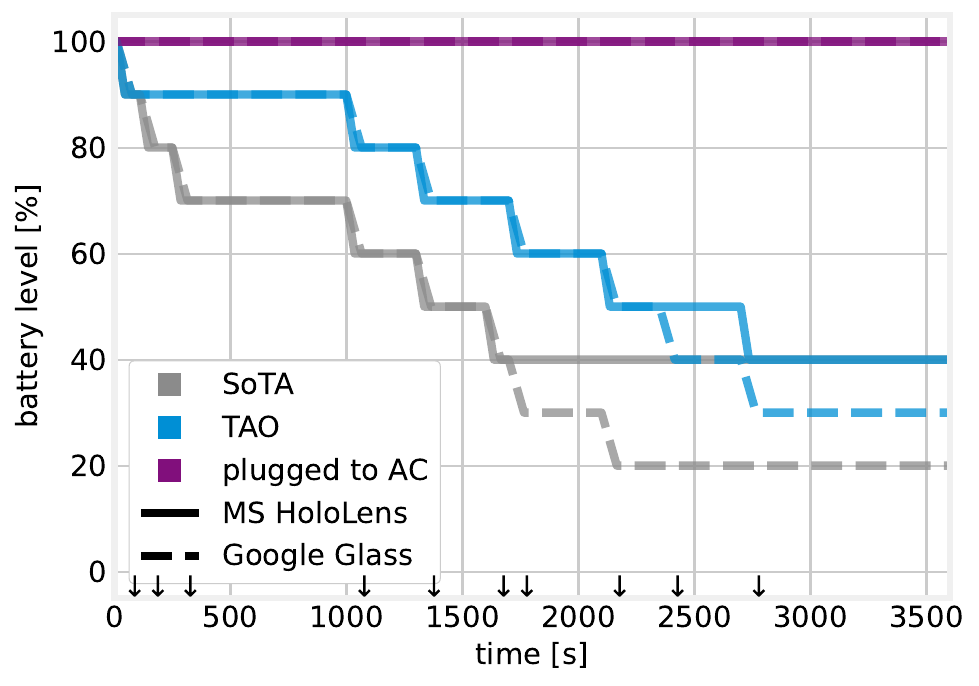}
\includegraphics[width=.30\textwidth]{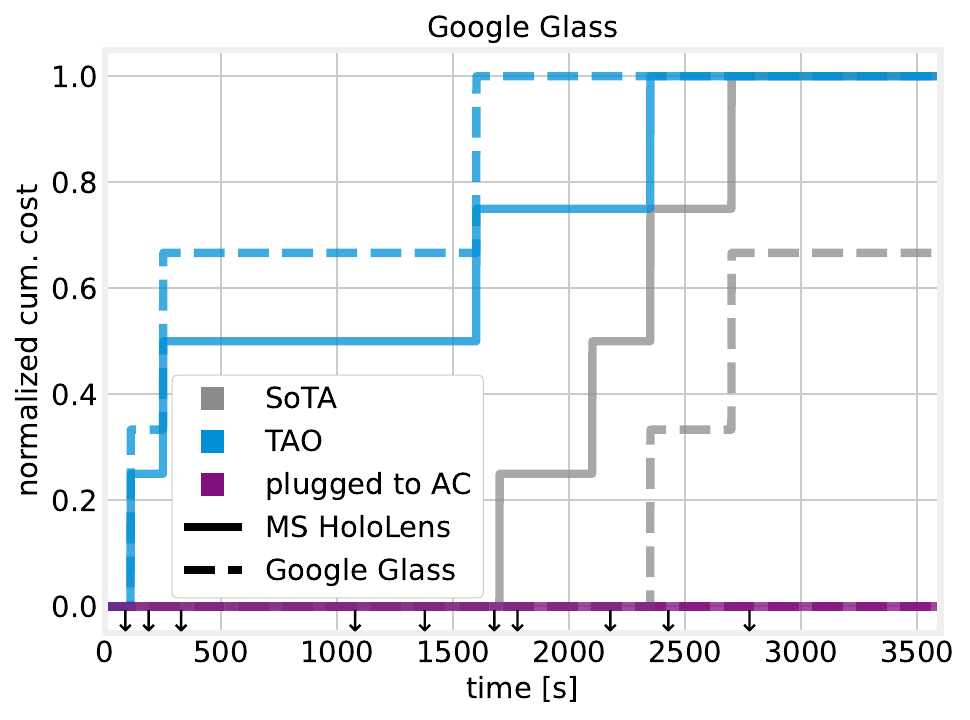}
\caption{
Empirical distribution of the wearable temperatures (left), time evolution of the battery level (center), and cost (right) yielded by each strategy.
\label{fig:peva2}
}
\end{figure*}

During the one-hour simulation period, a total of $10$~requests are generated at the times indicated by the black arrows on the bottom of the plots in Fig.~\ref{fig:peva1} and Fig.~\ref{fig:peva2} (center) and (right). For each request, the system must decide whether to process it locally or offload it. We compare three offloading strategies:
\begin{itemize}
    \item {\bf TAO}, solving the full problem formulated in Sec.~\ref{sec:deterministic} through the solution strategy presented in Sec.~\ref{sec:stationary};
    \item {\bf SoTA}, a state-of-the-art baseline that considers battery and power limits, but ignores the thermal constraint (\ref{eq:tau-max}).
    \item \textbf{Plugged to AC}, a greedy strategy that does not need to account for the battery duration, but only for the thermal design power. This represents the case of a wearable plugged into a wall socket.
\end{itemize}
For TAO, we set the confidence level to~$\omega=0.99$.

\begin{table}
\caption{
TDP and local processing time for evaluated wearables
    \label{tab:hw}
}
\begin{tabularx}{1\columnwidth}{|X||c|c|}
\hline
& Microsoft HoloLens & Google Glass\\
\hline\hline
TDP [W] & 2 & 0.6\\
\hline
duration of local processing [s] & 35 & 65\\
\hline
\end{tabularx}
\end{table}

Fig.~\ref{fig:peva1}(center) and Fig.~\ref{fig:peva1}(right) report the temperature evolution for the HoloLens and Google Glass, respectively. In both cases, TAO (blue lines) limits temperature spikes due to local processing by selectively offloading, while SoTa (grey lines) processes more requests locally, resulting in repeated temperature violations.
The effect is even more significant when the wearable is plugged to the AC (purple lines), as even more requests are served locally.

Local processing causes an immediate temperature rise, followed by a slower decrease after completion. In the first $500$~seconds of simulation during which three requests arrive, TAO processes only the first one locally (one spike), while the alternatives process all three, causing  the device to exceed the temperature limit (dotted red line). This is confirmed by Fig.~\ref{fig:peva2}(left), reporting the empirical temperature distribution: TAO (blue) consistently keeps temperature within safe limits, while SoTA (grey)
and ``plugged to AC'' (purple) exceed the thermal threshold, for about~5\% of the time in the former case, and even more in the latter case.
These results highlight the importance of thermal-aware decision-making in XR offloading strategies to avoid user experience degradation and health risks from overheated devices during prolonged skin contact.

Fig.~\ref{fig:peva2}(center) and Fig.~\ref{fig:peva2}(right) show, respectively, the time evolution of battery level and offloading cost. Consistent with Sec.~\ref{sec:deterministic}, local request processing depletes battery, causing the decline in battery level over time (downward trends in Fig.~\ref{fig:peva2}, center). Offloading requests to the edge increases costs (upward trends in Fig.~\ref{fig:peva2}, right). All strategies respect battery limits by enforcing constraint~\eqref{eq:b-min}, preventing battery depletion throughout the simulation. For Google Glass (dashed lines), the TAO strategy reduces local computation, resulting in higher remaining battery but increased offloading cost. This indicates that thermal constraints have a significant impact on the trade-off between local processing and edge offloading. For the HoloLens, the temperature constraint does \textit{not} noticeably reduce local computation or increase offloading cost. This suggests that accounting for thermal behavior can ensure safe operation without compromising performance or incurring additional costs.
Finally, if the wearable is plugged to the AC (purple line), the battery level is always~100\%, and no costs are incurred as all requests are served locally.

\begin{table}
\caption{
Cost incurred by TAO for different values of~$\omega$
    \label{tab:omegas}
}
\begin{tabularx}{1\columnwidth}{|X||c|c|c|c|c|}
\hline
Confidence level, $\omega$ & 0.80 & 0.85 & 0.90 & 0.95 & 0.99\\
\hline\hline
Normilized cost & 0.33 & 0.45 & 0.52 & 0.65 & 0.84\\
\hline
\end{tabularx}
\end{table}

We now study the effect of reducing the confidence level, all the way to~$\omega=0.8$. As summarized in Table~\ref{tab:omegas}, reducing~$\omega$ results in a significant lower cost, with savings exceeding~50\% for the lowest confidence levels. It is important to recall that there is no fixed, {\em right} confidence level: by allowing us to adapt such a level to the scenario at hand, TAO can yield   significant savings whenever lower confidence levels are acceptable, without losing its ability to tackle more critical scenarios with higher confidence levels.

\section{Conclusions and Future Directions}\label{sec:concl}

We investigated the role of XR wearables and highlighted the unique challenges they pose, namely, limited power and battery capacity, and susceptibility to overheating. To address these issues, we introduced TAO, a novel temperature-aware edge-based offloading strategy that considers both traditional metrics, such as battery life and power, and crucial thermal constraints. Our COMSOL-based simulations show that,
unlike its alternatives, TAO avoids exceeding temperature limits, while keeping additional edge offloading to a minimum.

Building upon our findings, several research avenues,  related to the XR-related tasks, can be pursued, including improved gesture recognition, voice synthesis, and hands-free control. In all these cases, user interfaces must adapt to various XR devices and screen dimensions, with UX/UI enhancements tailored to device temperature and environmental conditions. On the algorithmic side, one can exploit ML approaches to predict demand patterns and scale resources for dynamic workloads, integrated with temperature management strategies.

\bibliographystyle{ieeetr}
\bibliography{refs}

\newpage
\end{document}